\documentclass[preprint,5p,times,twocolumn]{elsarticle}

\date{}

\makeatletter
\def\ps@pprintTitle{%
 \let\@oddhead\@empty
 \let\@evenhead\@empty
 \def\@oddfoot{}%
 \let\@evenfoot\@oddfoot}
\makeatother

\usepackage{xpatch}
\xpatchcmd{\MaketitleBox}{\hrule}{}{}{}
\xpatchcmd{\MaketitleBox}{\hrule}{}{}{}

\usepackage{graphicx,lineno,xcolor,url,subfigure,float}
\usepackage{hyperref,lineno}
\usepackage{soul}
\hypersetup{colorlinks=true,citecolor=blue,urlcolor=blue}

\begin{document}

\begin{frontmatter}

\title{On the role of ion potential energy in low energy HiPIMS deposition: \\An atomistic simulation}

\author[1,2]{Movaffaq Kateb}
\author[2,3]{Jon Tomas Gudmundsson}
\author[4]{Pascal Brault}
\author[1]{Andrei Manolescu}
\author[2]{Snorri Ingvarsson}
\address[1]{Department of Engineering, School of Technology, Reykjavik University, Menntavegur 1, IS-102 Reykjavik, Iceland}
\address[2]{Science Institute, University of Iceland,
Dunhaga 3, IS-107 Reykjavik, Iceland}
\address[3]{Space and Plasma Physics, School of Electrical Engineering and Computer Science, \\KTH Royal Institute of Technology, SE-100 44, Stockholm, Sweden}
\address[4]{GREMI UMR7344 CNRS, Universit{\'e} d'Orl{\'e}ans, BP6744, 45067 Orleans CEDEX 2, France}

\begin{abstract}
We study the effect of the so-called \emph{ion potential} or non-kinetic energies of bombarding ions during ionized physical vapor deposition of Cu using molecular dynamics simulations. In particular we focus on low energy high power impulse magnetron sputtering (HiPIMS) deposition, in which the potential energy of ions can be comparable to their kinetic energy. The ion potential, as a short-ranged repulsive force between the ions of the film-forming material and the surface atoms (substrate and later deposited film), is defined by the Ziegler-Biersack-Littmark potential. Analyzing the final structure indicates that, including the ion potential leads to a slightly lower interface mixing and fewer point defects (such as vacancies and interstitials), but resputtering and twinning have increased slightly. However, by including the ion potential the collision pattern changes. We also observed temporary formation of a ripple/pore with 5~nm height when the ion potential is included. The latter effect can explain the pores that have been observed experimentally in HiPIMS deposited Cu thin films  by atomic force microscopy.
\end{abstract}


\end{frontmatter}


\section{Introduction}
Over the past few decades, high power impulse magnetron sputtering (HiPIMS) has attracted significant attention among the variety of ionized physical vapor deposition (PVD) techniques \citep{helmersson06:1,gudmundsson12:030801}. In dc magnetron sputtering (dcMS) low plasma density (10$^{15}$\,--\,10$^{17}$~m$^{\,-3}$ \citep{tummi2020}) and low ionization  fraction in the deposition flux ($<$10\% \citep{christou00:2897}) is dictated by the thermal load at high powers, which might be destructive to the cathode target. As a result, the majority of the film forming species arriving at the substrate surface in dcMS are electrically neutral and the majority of the ions arriving at the substrate are ions of the working gas. In HiPIMS the overheating issue of the cathode target is resolved by applying unipolar high power pulses with a low duty cycle and at low repetition frequency, while maintaining similar time averaged power as for dcMS \citep{helmersson06:1,gudmundsson12:030801}. Therefore a peak electron density that is 2\,--\,3 orders of magnitude higher than in dcMS, is achieved in the vicinity of the cathode target \citep{gudmundsson02:249,bohlmark05:346,meier18:035006}. Due to the high electron density the ionization mean free path of the sputtered species becomes shorter and the ionization probability increases and a significant fraction of the ions reaching the substrate are ions of the film-forming species. As a result HiPIMS deposition presents smoother \citep{magnus11:1621,sarakinos07:2108,kateb2019}, denser \citep{samuelsson10:591,kateb2019}, and less defective \citep{alami05:278,kateb2019} coatings, compared to dcMS deposition. Further, by applying a substrate bias, the bombarding energy of the ions of the film forming material can be tuned in order to achieve desired film properties such as the film texture and grain size \cite{hajihoseini2018,kateb2020}.

Molecular dynamics (MD) simulations have been shown to be a promising tool in terms of revealing atomistic mechanisms that contribute to various PVD techniques including dcMS and HiPIMS \citep{kateb2019,kateb2020}. It is worth mentioning that MD simulations of sputter deposition are divided into two categories: (i) modeling the vicinity of the substrate, the so called \emph{deposition side}, and (ii) modeling the energetic working gas ions bombarding a surface or the \emph{target side}. So far the target side simulations are focused on the sputter yield and energy distribution of ejected atoms from the target \citep{brault2016,kammara2016}. Recently, \citet{brault2018} included the working gas between target and deposition sides within a multi-scale MD simulation. He suggested that the effect of the gas phase in sputter deposition is limited to the narrowing of the energy distribution as atoms travel towards the substrate. It has also been shown that one can model the kinetic energy of adatoms at desired pressure without considering a working gas in the simulation \citep{xie2014}. Thus, it is reasonable to only model the deposition side using a proper energy distribution within the deposition flux.

Regarding HiPIMS deposition, MD simulations have indicated that both ionization fraction and substrate bias change the microstructure and film-substrate intermixing \citep{kateb2019,kateb2020}. Hence, the variation of the film texture by bias voltage that has been observed experimentally \citep{cemin17:120,nedfors18:031510,hajihoseini2018} can be attributed to the change of wetting due to interface mixing. An alternative explanation is based on  \emph{thermal spikes} \citep{johnson1985} or the \emph{atomic scale heating} \citep{musil2000} models of impinging ions.
However, an MD simulation in the absence of ions is inconsistent with these models, as pointed out by  \citet{gilmore1991}. Even energetic deposition (10\,--\,40~eV) failed to make connection between MD results and the above mentioned models \citep{sprague1996}. On the other hand, \citet{muller1986} showed that low energy ion impacts (ions of the noble working gas) can generate thermal spikes that lead to structure modification. Later, \citet{anders02:1100} explained that the non-kinetic energy of ions, which will be called \emph{ion potential} here after, cannot be ignored, and introduced several energy terms of an ion: 
\begin{equation}
    {\cal E}_{\rm i}={\cal E}_{\rm i}^0+ne\Delta V+{\cal E}_{\rm c}+{\cal E}_{\rm iq}+{\cal E}_{\rm exc}+\sum_{Q=1}^{n} {\cal E}_{Q} \ ,
    \label{eq:anders}
\end{equation}
where ${\cal E}_{\rm i}^0$ is the initial kinetic energy of the ion of the sputtered species, $n$ is the ion valance, $\Delta V$ is the potential difference between the substrate and the plasma,  ${\cal E}_{\rm c}$ is the cohesive energy, ${\cal E}_{\rm iq}$ is the image charge interaction energy, and ${\cal E}_{\rm exc}$ is a contribution  that only applies to ions with excited electron(s). Finally, ${\cal E}_{Q}$ is a cumulative ionization energy i.e.~the summation of the first, second, ... and up to the $n$-th ionization energy to be considered for multiply charged ions. 
Note that the terms ${\cal E}_{\rm i}^0$ and ${\cal E}_{\rm c}$  apply both to neutral atoms and ions.  The term $ne\Delta V$ takes into account the possibility to control an ionized flux using a substrate bias. Terms ${\cal E}_{\rm c}$, ${\cal E}_{\rm exc}$, ${\cal E}_{Q}$ define the ion potential indicated by \citet{anders02:1100,anders2008}, which must be treated carefully in order to achieve a realistic simulation of an ionized PVD.

In an MD simulation ${\cal E}_{\rm i}^0$ can be defined in terms of initial velocity. Since the simulated vacuum near the substrate is smaller than the substrate sheath ($<$\,1~mm), it is not possible to model acceleration/deceleration of ions due to the substrate bias. Instead, it is convenient to consider the initial velocity corresponding to the summation of ${\cal E}_{\rm i}^0$ and   $ne\Delta V$ \citep{kateb2020}. Since we are modelling the vicinity of the substrate, the above mentioned initial velocity refers to the velocity prior to impact. 

The term ${\cal E}_{\rm c}$ is normally associated with the interatomic potential i.e.\ an individual atom in the vacuum has zero potential while the potential of an atom in the bulk is equal to ${\cal E}_{\rm c}$. The image charge term is caused by the attraction between ions and image charges (electrons) that appears within a few nanometer depth from the surface \citep{anders2008}.  
According to the \emph{over-barrier} model \citep{burgdorfer1991}, for a low valence ion that approaches a metallic surface, neutralization occurs at a critical distance from the surface. This allows ignoring the ${\cal E}_{\rm iq}$ during deposition on a metallic surface \citep{kateb2020}. Both ${\cal E}_{\rm exc}$ and ${\cal E}_{\rm Q}$ can be taken into account using the electron force field (EFF) \citep{su2007} which is available for $s$ and $p$ electrons \citep{jaramillo2011}. However, utilizing EFF is computationally intensive and increases the system size for taking into account electrons. 


An interesting possibility is modeling the ion potential as an extra short-ranged repulsive interatomic potential \citep{byggmastar2018,kateb2019,kateb2020}. The term extra means that we already included an interatomic potential due to ${\cal E}_{\rm c}$ that might accurately describe forces around the equilibrium interatomic distances but not short-ranged ones \citep{byggmastar2018}.
This method was first utilized by \citet{muller1986} to model working gas ions. He showed that bombarding a growing film by ions of the working gas increases film density and smoothness \citep{muller1987prb,muller1987}. 
Also, a higher ratio of working gas ions to neutral atoms was found to reduce the number of voids \citep{muller1986}. Later, \citet{fang1993} applied this model to the film forming ions and obtained a negligible effect from the working gas ions. Recently, \citet{byggmastar2018} applied the model for a realistic simulation of cascade damage. They showed that the Ziegler-Biersack-Littmark (ZBL) potential \citep[Chap.~2]{ziegler1985} can compensate the underestimation of the repulsive range by the embedded atom method (EAM) \citep{daw1983,daw1984}. As a result, the threshold displacement energy and many-body repulsion at short interatomic distances can be precisely modeled. 

More recently, we utilized a combination of the ZBL potential and the EAM for a simulation of deposition with partially and fully ionized flux of sputtered species \citep{kateb2019}. It has been shown experimentally that surface roughness decreases as higher ionization fraction is utilized  \citep{magnus11:1621,sarakinos07:2108}. This occurs through the so-called \emph{bi-collision event}, that causes a localized amorphization followed by recrystallization. The latter was not observed in the previous MD simulation studies without considering ions \citep{houska2014,xie2014}. We could also observe in the simulation results the formation of twinning during epitaxial growth of Cu using HiPIMS \citep{kateb2019}, that had been reported in previous experiments \citep{cemin2017,kateb2019epi}. Then, we showed that tuning the kinetic energy of ions through a substrate bias allows minimizing the creation of defects \citep{kateb2020}. A similar trend of polycrystalline to epitaxial transition using a substrate bias has also been observed experimentally \citep{cemin17:120}. Moreover, it has been shown that energetic deposition (e.g.\ in highly biased HiPIMS or cathodic arc deposition) causes re-sputtering of deposited film and substrate, a process known as \emph{potential sputtering} \citep{winter2001}. This effect explains the reduction of Cu deposition rate with increased substrate bias observed experimentally \citep{cemin17:120}.

The above mentioned examples clearly indicate the usefulness and reliability of describing the ion potentials using the ZBL term. The aim of the present manuscript is to study the effect of the ionic potential itself during the bombardment of a solid. We compare two cases, with and without considering the ZBL potential between ion and surface (substrate and later deposited film), and discuss changes in the film properties. We consider copper as the material of choice and assume a fully ionized flux of sputtered species to represent the HiPIMS process.  This assumption is based on experimental findings from HiPIMS discharges with Cu target. For a Cu target, during HiPIMS operation the ionization fraction of the sputtered flux up to 70\% has been reported \citep{kouznetsov99:290}. We can also ignore the working gas ions because: (i) using HiPIMS, the working gas and metal ions arrive at the substrate at different times \citep{macak00:1533} and (ii) in the case of Cu it has been reported that up to 92\% of ions arriving at the substrate are ions of the target material \citep{vlcek07:45002}. We neglect the contribution of ${\cal E}_{\rm exc}$, as there is a low probability of finding excited ions far from the target \citep{anders2008}.
For Cu the first ionization energy is also relatively low (7.73~eV), and it can also be neglected compared to the ion kinetic energies \citep{anders02:1100}.

\section{Method}

MD simulations \citep{allen1989} were performed using the LAMMPS \citep{plimpton1995,plimpton2012} package\footnote{LAMMPS website, \url{http://lammps.sandia.gov/}, distribution 14-April-2018}. The EAM force field was employed to model the interactions of film/substrate atoms. In the EAM, the total potential energy of atom $i$ (${\cal E}_i$) is described by \citep{daw1983,daw1984}
\begin{equation}
	{\cal E}_i=F_i(\rho_i)+\frac{1}{2}\sum_{i\neq j}\phi_{ij}(r_{ij}) \ ,
\end{equation}
where $F_i$ is the embedding energy of atom $i$ into electron density $\rho_i$ and $\phi_{ij}$ is a pair potential interaction of atom $i$ and $j$ at distance $r_{ij}$. Note that the electron density $\rho_i$ itself depends on electron density of neighboring atoms $\rho_{ij}$ so that $\rho_{i}=\sum_{j}\rho_{ij}(r_{ij})$.

%

The EAM potential relies on the same principles as the density functional theory and has been successfully applied to determine equilibrium and near-equilibrium behaviour of various metals and alloys. However, energetic deposition might push the system far away from its equilibrium. To describe the short-ranged forces acting when an energetic atom travels inside the lattice we utilized the ZBL potential that has already been implemented in many force fields such as EAM \cite{kammara2016,byggmastar2018}, Tersoff \cite{bellido2012} and deep-learning \cite{wang2019} potentials. The ZBL potential is a modification of screened Coulomb potential \citep[Chap.~2]{ziegler1985}
\begin{equation}
	V(r_{ij})=\frac{Z_iZ_je^2}{4\pi\varepsilon_0r_{ij}}\Phi\left(\frac{r_{ij}}{a}\right) \ ,
\end{equation}
where the $Z_i$ and $Z_j$ are the atomic numbers of species $i$ and $j$ that belong to the Coulomb term and $e$ and $\varepsilon_0$ stand for elementary charge and vacuum permitivity, respectively. Note that the Coulomb term includes the atomic numbers in order to take into account very short interatomic distances ($<$2~{\AA}) where electron-electron repulsion is dominating. Various ion potentials, such as Neilson, Brinkman and Firsove \citep[Chap.~2]{townsend1976}, consider a similar Coulomb term while their screening function ($\Phi$) might change. The term $\Phi$ was introduced by \citet{bohr1948} and later modified by several authors. Finally, we apply  the universal screening function $\Phi$ of   \citet{ziegler1985} as discussed in our earlier work \citep{kateb2019,kateb2020}.
%
%
%
%
%
The interested reader is referred to \citet[Chap.~2]{townsend1976} for comparison of various ion potential descriptions for the case of Cu. 

We considered two scenarios of ion-surface (substrate, and later deposited film) interaction: (i) EAM/ZBL and (ii) only EAM i.e.\ with and without ion potential, respectively. The repulsive nature of the ZBL potential may also result in exaggerated etching at high deposition rate. This issue can be solved by a checking mechanism, i.e.\ if an ion stays on the surface or implants into sublayers and remains there for 1~ps, it is considered to be an atom belonging to the film and its ZBL potential with respect to film atoms will be turned off. This 1~ps is consistent with the time scale for the electrons and the lattice to reach an equilibrium state after the localized heating due to high energy ion impacts \cite{anders02:1100}.

It is also worth mentioning that the ion-ion interaction in the flux was modeled via a hybrid approach based on EAM and ZBL potentials. This might seem unnecessary since the flux travel towards the substrate is conventionally assumed to be collision-less. The latter is only true within a fixed energy flux, otherwise the energy distribution allows gas phase collisions \citep{kateb2019,kateb2020}. Recently, \citet{brault2018} demonstrated the gas phase clustering using a multi-scale MD simulation, i.e. when the deposition flux travels a distance similar to that in a typical experiment, with gas phase collisions treated properly.

The substrate was assumed to be a single crystal Cu, consisting of 16 (111) planes with 77$\times$90~{\AA}$^2$ dimensions. This makes the $\langle111\rangle$ orientation to be the growth direction. The substrate was divided into 3-layers as proposed by \citet{srivastava1989}. A fixed monolayer is set at the bottom to prevent the other layers from moving after collisions, then three monolayers are set as a thermostat layer to control the heat dissipation and prevent melting, and the rest of the substrate is the surface layer. The initial velocities of substrate atoms were defined randomly from a Gaussian distribution, to mimic a temperature of 300~K, and the substrate energy was minimized prior to relaxation.

For both cases studied, the deposition flux consisting of 22000 ions was introduced at a distance of 150~nm above the substrate surface. The initial velocities of the ions towards the substrate were assigned randomly, with a flat distribution corresponding to the 0.1\,--\,40~eV energy range. The flat energy distribution may seem a rather crude  approximation. The energy distribution within deposition flux has been measured experimentally for both dcMS and HiPIMS.  The original measurement of the ion energy distribution for Ti$^+$ ions in HiPIMS operation indicates a flat distribution up to 40~eV which then levels off   \citep{bohlmark06:1522}, while for Cu$^+$ it levels off around 20 eV \citep{vlcek07:45002}. 
The process of introducing species was a single ion every 0.1~ps, which produces an equal deposition rate in both cases. One may think this will generate deposition rate several orders of magnitude larger than in a typical experiment. We would like to remark that, considering the pulse duration, the instantaneous deposition rate of HiPIMS is 1\,--\,2 orders of magnitude higher than for dcMS  \citep{kateb2018}. Besides, experimental measurements indicate a variable intensity of the ion flux during the pulse, reaching its peak within a fraction of the pulse \citep{greczynski19:060801}. Therefore, for HiPIMS the instantaneous deposition rate is typically several orders of magnitude higher than that of e.g.\ dcMS. 
However, the time scale that is achieved in MD simulation is generally on the order of tens of ns. Thus, MD simulations cannot capture the entire HiPIMS pulse which is normally 50\,--\,400 $\mu$s long \citep{gudmundsson12:030801}.

The time integration of the equation of motion was performed using the Verlet algorithm \citep{verlet1967,kateb2012} with a timestep of 5~fs and sampling from the microcanonical (NVE) ensemble. The Langevin thermostat \citep{schneider1978} was only applied to the thermostat layer with a damping of 5~ps for a total time of 2.5~ns. The damping defines the timescale for resetting the temperature and that generates a heat flow towards thermostat layer.

In order to study the film-substrate interface quantitatively, we used the partial radial distribution function, $g_{ij}(r)$ introduced by \citet{ashcroft1967}, that was originally proposed for homogeneous binary mixtures. Here we have adapted it to the case of interface mixing, with the assumptions explained elsewhere \citep{kateb2020}.

We considered the common neighbor analysis (CNA) method \citep{steinhardt1983,kelchner1998,tsuzuki2007} to characterize the local structure. CNA allows us to distinguish between face centered cubic (fcc) and hexagonal close-packed (hcp) structures which is of practical importance in defect analysis \citep{kateb2018b,azadeh2019}. This is achieved by determining the angles between the nearest neighbors, and thus the fcc and hcp lattices, having central and mirror symmetry in (111) planes, respectively, can be distinguished by a slight angle difference. The Ovito package\footnote{Ovito website, \url{http://ovito.org/}, Version 3.0.0-dev794} was used to generate atomistic illustrations and its Python interface was used for post-processing of CNA and $g_{ij}(r)$ \citep{stukowski2009}. It is worth mentioning that Ovito allows utilizing an adaptive CNA, demonstrated by  \citet{stukowski2012,stukowski2014} with variable cutoff calculation. This means that it is not sensitive to the choice of cutoff for finding the nearest neighbors. However, even the adaptive CNA characterizes atoms located at the surface as disordered, due to the lack of symmetry.

Furthermore, we studied vacancy and interstitial defects using the Wigner-Seitz (WS) method \citep{nordlund1998}, based on the comparison of a current structure with an initial or a perfect lattice as reference. The method generates a primitive cell volume for each atom in the reference lattice and counts current atoms inside each volume. Thus, empty and over-occupied volumes translate to vacancy and interstitial defects.

\section{Results}



\subsection{Intermixing}
\begin{figure}
    \centering
    \includegraphics[width=1\linewidth]{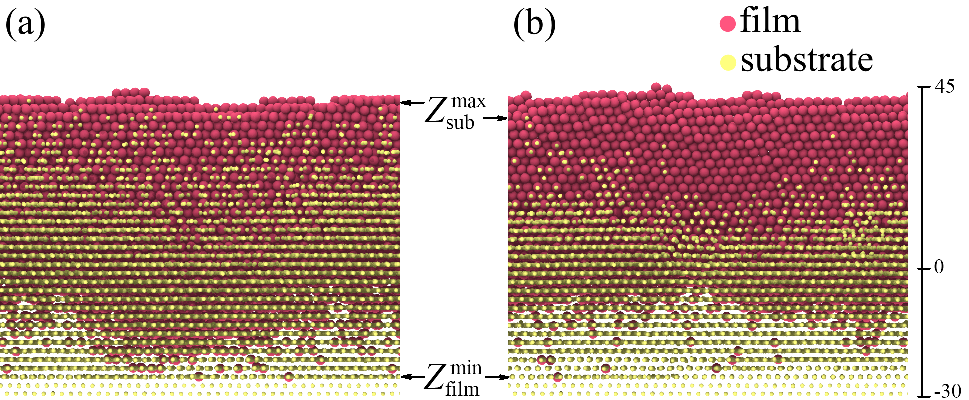}
    \caption{Cross-section of the film/substrate layers (a) with and (b) without considering ion potential. The film atoms are illustrated by larger red atoms, and substrate atoms are depicted smaller yellow atoms, for clarity. Note that the substrate atoms are shifted in front of film atoms for a better illustration of the transition region. The scale on the right shows height from the initial substrate surface.}
    \label{fig:mix}
\end{figure}

Figure~\ref{fig:mix} shows the cross-section of the film/substrate layers, obtained with and without considering the ion potential in the simulation. In both cases, severe intermixing is obtained which is a characteristic of low biased HiPIMS deposition process \citep{kateb2020}. However, a relatively thicker transition region is obtained when the ion potential is included in the simulation. Table~\ref{tab:z} summarizes the results quantitatively. In both cases almost a 40~{\AA} thick film is deposited. The exact values are presented by $Z_{\rm film}^{\rm max}$ in Table~\ref{tab:z}. There is a deviation between the number of deposited species (22000) and the number of film atoms, $N_{\rm film}$. There are also losses in the substrate atoms indicated by $\Delta N_{\rm sub}$. These quantities indicate a slightly increased resputtering when considering the ion potential. This difference is referred to as potential sputtering effect \citep{winter2001}. The value of $Z_{\rm film}^{\rm min}$ refers to depth of the lowermost film atom which is about 26~{\AA} below the initial substrate surface for both cases.  We also located the topmost substrate atom in the film at the position denoted by $Z_{\rm sub}^{\rm max}$, which is at about 4~nm distance from original substrate surface. Assuming $Z_{\rm sub}^{\rm max}-Z_{\rm film}^{\rm min}$ represents the interface thickness, a value of $\sim$60~{\AA} is obtained with a negligible difference i.e.~it is 4~{\AA} larger with ion potential included in the simulation.

\begin{table}
\centering
\caption{\label{tab:z} $N_{\rm film}$ denotes number of film atoms and $\Delta N_{\rm sub}$ refers to sputtered substrate atoms, and $Z$ indicate maximum/minimum position of the film/substrate atoms with respect to the initial substrate surface.}
\begin{tabular}{ c c c }
\hline \hline
Method & with & without \\
\hline
$N_{\rm film}$ (atom) & 21646 & 21733 \\
$Z_{\rm film}^{\rm min}$ ({\AA}) & -25.8 & -25.9\\
$Z_{\rm film}^{\rm max}$ ({\AA}) & 42.8 & 44.0\\
$\Delta N_{\rm sub}$ (atom) & 202 & 147 \\
$Z_{\rm sub}^{\rm max}$ ({\AA}) & 40.8 & 36.6 \\
\hline \hline
\end{tabular}
\end{table}

\begin{figure}
    \centering
    \includegraphics[width=1\linewidth]{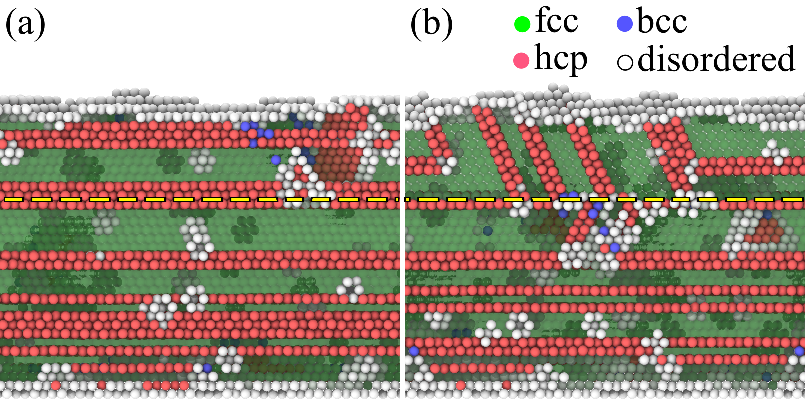}
    \caption{Demonstration of the microstructure using CNA for deposition (a) with and (b) without ion potential. The yellow dashed line indicate initial substrate surface. The fcc atoms are shown as semitransparent for illustration of internal defects.}
    \label{fig:cna}
\end{figure}

\begin{table}
\centering
\caption{\label{tab:defect} The results from the WS analysis of the final structure with $N^{\rm V}$ and $N^{\rm I}$ being the number of vacancies and interstitial atoms, respectively.}
\begin{tabular}{ c c c }
\hline \hline
Method & with & without \\
\hline
$N^{\rm V}_{\rm sub}$ (atom) & 911 & 959 \\
$N^{\rm I}_{\rm sub}$ (atom) & 427 & 448 \\
$N^{\rm V}_{\rm film}$ (atom) & 1066 & 1268 \\
$N^{\rm I}_{\rm film}$ (atom) & 1225 & 1288 \\
\hline \hline
\end{tabular}
\end{table}

\subsection{Microstructure}
The CNA analysis of films deposited with and without  considering the ion potential is shown in figure~\ref{fig:cna}. Various defects can be observed and characterized in both cases. The presence of a hcp layer in a fcc material is generally referred to as a stacking fault, or a twinning when it is large enough. Vacancies can be detected using CNA as empty sites enclosed by 12 disordered atoms. A tetrahedral or octahedral configuration of disordered atoms also represent an interstitial. Further details on the point defects analysis by WS are summarized in Table~\ref{tab:defect}. The dashed line in figure~\ref{fig:cna} indicates the original substrate surface that is used to distinguish between the film and substrate defects. 

It can be seen that, in both cases, low energy HiPIMS deposition induces a considerable number of various defects into the substrate. Introducing defects into the substrate is the characteristic of ionized PVD and does not occur during e.g.\ evaporation when almost all the film-forming species are neutral \citep{kateb2019,Kateb2020b}. However, as previously demonstrated \citep{kateb2020}, high ionization fraction of HiPIMS allows us to tune the kinetic energy of the flux using the substrate bias and efficiently minimizing the defects. WS analysis also indicates a considerable number of point defects, both in the film and in the substrate. Note that, however, the density of vacancies is more uniform in the film and substrate, while interstitials are mostly located in the films. This means that both films here present a more compressive stress than their substrates.

Regarding the effect of ion potential on the microstructure, it can be seen that without the ion potential more point defects are introduced both in the film and in the substrates. However, the ratio of hcp atoms is higher when the ion potential is taken into account.

\subsection{Temporal behaviour}
\begin{figure}
    \centering
    \includegraphics[width=1\linewidth]{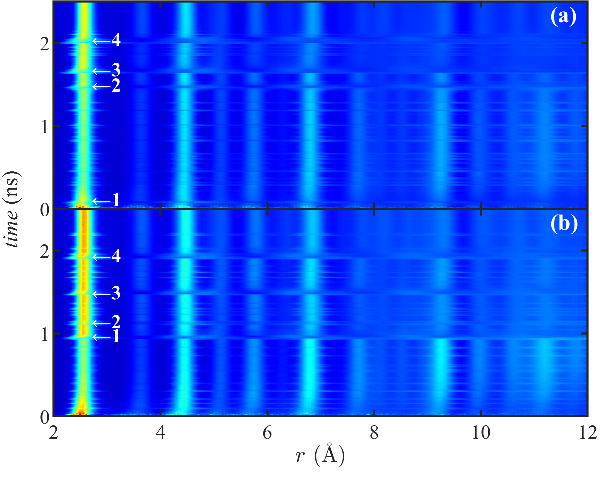}
    \caption{Variation of $g_{ij}(r)$ during the deposition (a) with and (b) without ion potential. The colormap indicates normalized density of film-substrate atom pairs. Arrows indicate considerable shift in 1NN peaks due to major collisions.}
    \label{fig:Gr}
\end{figure}

Figure~\ref{fig:Gr} illustrates $g_{ij}(r)$ as a function of the elapsed time during the simulated deposition with and without the ion potential. The color map indicates the intensity of the normalized $g_{ij}(r)$ i.e.\ the density of film-substrate bonds. The major peak indicates the distance between the first nearest neighbors (1NN). That is 2.49~{\AA} at the early stage of deposition, and becomes 2.55~{\AA} with the formation of the film. Such an increase in the 1NN generally indicates an increase in the coordination number, or, here,  per atom film-substrate bonds. The disturbances in the $g_{ij}(r)$, which are more evident in the major peaks, mean the redistribution of atoms due to collisions. Each collision shifts the 1NN peak to lower distances, that indicates a compression of the system. Then, the 1NN peaks relax to their original position. The arrows point to four major collisions that cause a clear shift in the 1NN peak. For a flat interface it is expected that $g_{ij}(r)$ tends to zero as the film thickness increases. This is because in a flat interface the number of film-substrate bonds remains constant while the total number of bonds grows. In both cases here, the intensity of the major peak in general increases. This is an indication of a continuous interface mixing and the development of more film-substrate bonds.

\begin{figure}
    \centering
    \includegraphics[width=1\linewidth]{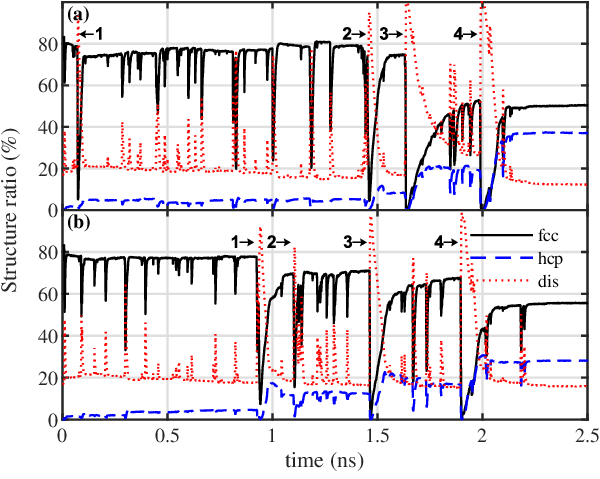}
    \caption{Evolution of microstructure obtained by CNA for deposition (a) with and (b) without ion potential. The arrows correspond to the collisions indicated in figure~\ref{fig:Gr}.}
    \label{fig:cna-time}
\end{figure}

Next, figure~\ref{fig:cna-time} shows the time evolution of the microstructure during deposition. The arrows here correspond to collisions indicated in figure~\ref{fig:Gr}. Looking into the fcc percentage, it can be seen that after the indicated collisions the system requires some time for recovery. The only exception is collision 1 in figure~\ref{fig:cna-time}(a), that occurs at the early stage of deposition. The collisions indicated by arrows produce a severe amorphization followed by a step change in the hcp percentage. We observed that the rest of the collisions can only produce point defects. To our knowledge \citet{klokholm1969} was the first to suggest the existence of a disordered liquid-like material buried beneath the advancing film surface during evaporation. He believed that, during the evaporation, the energy of adatoms (0.1\,--\,4~eV) is not high enough to cause the recovery of sub-surface layers. Instead, we used ions and much higher energies (0.1\,--\,40~eV), that allow for amorphous to crystalline phase change. However, even this range of energies is not sufficient for the formation of equilibrium fcc structure to proceed without defects. The latter can be achieved only at higher energies, like 60\,--\,100~eV \citep{kateb2020}.

We would like to recall that we used similar initial velocities in both cases studied. We found that when the ion potential is considered the collision pattern changes. Note that without the ion potential (see  figure~\ref{fig:cna-time}(b)) the collisions indicated by arrows are more regularly spaced compared to the case with the ion potential included in the simulation. This is mainly because the presence or the absence of the ionic potential leads to a variation of the kinetic energy in vicinity of the surface.

\begin{figure}
    \centering
    \includegraphics[width=1\linewidth]{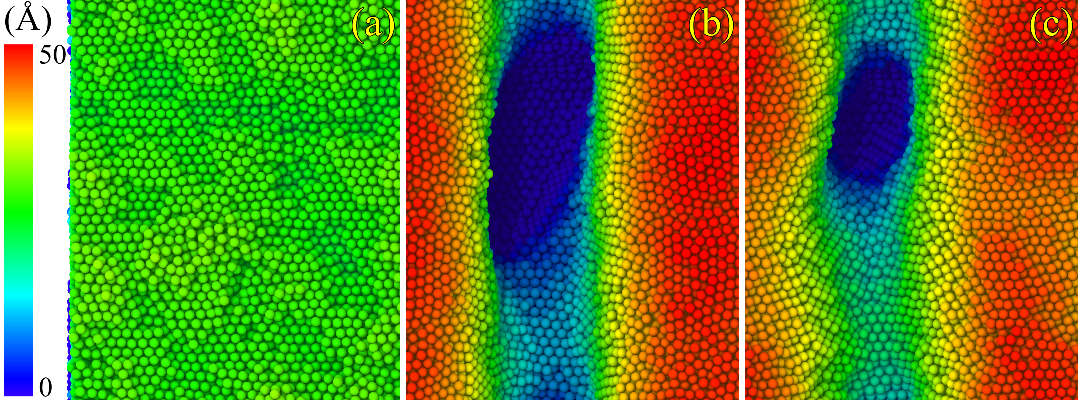}
    \caption{Temporary surface roughness due to the ion potential obtained (a) before and (b) after collision 3 (figure~\ref{fig:cna-time}(a)), and (c) before collision 4.}
    \label{fig:z}
\end{figure}

Figure~\ref{fig:z} shows a temporary surface roughness that appears after collision 3 in figure~\ref{fig:cna-time} (a). This ripple/pore remains almost unchanged until collision 4. The formation of similar ripples, with significant surface roughness, has been observed after ion irradiation of Si surface under an angle using kinetic Monte Carlo (KMC) simulations within binary collision approximation \citep{liedke2013}, hybrid MD/KMC \citep{Yang2015}, and also experimentally   \citep{madi2008,Keller2008}. While these studies were focused on ions incident under an angle, e.g. 80$^\circ$, we considered the initial velocity of flux perpendicular to the substrate surface. However, the appearance of a pore indicates an elongation that is a characteristic of oblique incidence. Thus, the deflection of the energetic ions was achieved due to the repulsion between the ions and the surface. We did not observe the formation of any pore without considering the ion potential. This is due to the fact that considering ion potential would lead to a more realistic collision behaviour, such as possibility of deflection, while the average behaviour or overall behaviour for longer timescale may seem similar (cf.~figures.~\ref{fig:mix}--\ref{fig:cna-time}). We would like to remark that the formation of open pores has been verified by atomic force microscopy of Cu thin films deposited by HiPIMS \citep{cemin17:120}. To the best of our knowledge, in this work we have shown for the first time that such effects can occur due to the ionic potential. The formation of open pores at the surface has also been reported for relatively harder materials, such as CrN and NbN deposited by HiPIMS \citep{biswas2018}. Thus, our findings are of general interest for variety of thin film deposition processes.

It is worth mentioning that earlier we reported enhanced surface roughness obtained by utilizing a flux of neutrals \citep{kateb2019,kateb2020}. Such a surface roughening were caused by so-called \emph{steering effect} \citep{vandijken1999}. The latter causes surface roughness by preferential landing of arriving neutrals on top of islands due to the long-ranged attractive forces.  However, the ion potential here was applied through the ZBL potential which is of short-ranged repulsive nature. Besides, the pore shown in figure~\ref{fig:z} of the present manuscript, appears immediately after collision~3 (figure~\ref{fig:cna-time}(a)) and disappears after collision~4 while surface roughnesses observed using a flux of neutrals were formed gradually.

\subsection{Discussion}
A question that may rise here is when is it necessary to include ion potential? To answer the question we need to highlight couple of considerations. As mentioned earlier, most of the interatomic potentials, such as EAM are obtained by fitting some particular properties at equilibrium condition. Thus, low energy collisions, e.g.\ during deposition of neutral adatoms or partially ionized flux, that does not push the system out of equilibrium can be described without including ion potential. For high ionization flux fractions, the probability of collisions that may go far from equilibrium increases and one may consider including ion potential. 
This also depends on the ions kinetic energy i.e.\ very high kinetic energies deposit much more energy into the film which then makes the influence of the ion potential negligible. On the other hand, in the low kinetic energy regime, there is lower probability of collisions with non-equilibrium behaviour. However, there are conditions where ion potential must be included such as when  modelling the target side.

It is also worth mentioning that, conventionally ionized PVD methods have been modeled with energetic neutrals rather than ions (cf. Ref. \citep{kateb2019} and references there in). Now another question arises: can energetic neutrals compensate for the ion potential? The answer is that the ion potential is effective in a very short interatomic distance that does not occur in all collisions. Thus, the effect of the ion potential is very localized. However, energetic neutrals bombard everywhere on the surface and modify the structure more homogeneously.


\section{Summary}
In summary, the effect of the ion potential energy on a film deposited by highly ionized flux of the film-forming material, such as during HiPIMS deposition is studied using MD simulations. We considered two cases, with and without considering the ZBL repulsive potential between ions and surface. We have shown that there is a slight difference between the two cases, in terms of surface roughness, interface mixing, point defects, the microstructure, and the time evolution. In addition, we have shown that the ion potential changes the collision pattern. In particular, we have obtained a temporal ripple/pore with 50~{\AA} height difference, that can explain experimental results. 

\section*{Acknowledgements}
This work was supported by the University of Iceland Research Funds and the Icelandic Research Fund, grants nos.\ 195943, 196141, 130029 and 120002023.


\bibliography{ref}

\end{document}